\documentclass[bm,aps,showpacs,amsfonts,amssymb,preprint,nofootinbib]{revtex4}

\usepackage{graphicx}
\usepackage{natbib}
\usepackage{amsmath}

\usepackage{color}
\font\mybb=msbm10 at 12pt

\font\mybbsub=msbm10 at 8pt

\def\bb#1{\hbox{\mybb#1}}

\def\bbsub#1{\hbox{\mybbsub#1}}

\def\RR {\bb{R}}

\def\RRsub {\bbsub{R}}

\def\CC {\bb{C}}
\def\CCsub {\bbsub{C}}
\def\HH {\bb{H}}

\def\AA {\bb{A}}
\def\AAsub {\bbsub{A}}

\def\OO {\bb{O}}

\newcommand\beqa{\begin{eqnarray}}
\newcommand\eeqa{\end{eqnarray}}
\newcommand\n{\nonumber\\}
\begin{document}

{~}

\title{On an Algebraic Structure of Dimensionally Reduced \\Magical Supergravity Theories}
\vspace{2cm}
\author{Shin Fukuchi\footnote[1]{E-mail:fshin@post.kek.jp} and Shun'ya Mizoguchi\footnote[2]{E-mail:mizoguch@post.kek.jp}}
\vspace{2cm}
\affiliation{\footnotemark[1]\footnotemark[2]SOKENDAI (The Graduate University for Advanced Studies)\\
Tsukuba, Ibaraki, 305-0801, Japan 
}
\affiliation{\footnotemark[2]Theory Center, Institute of Particle and Nuclear Studies,
KEK\\ Tsukuba, Ibaraki, 305-0801, Japan 
}
\begin{abstract} 
We study an algebraic structure of magical supergravities 
in three dimensions. 
We show that 
if the commutation relations among the generators 
 of the quasi-conformal group in the super-Ehlers decomposition
are in a particular form, then 
one can always find a parameterization of the group element 
in terms of various 3d bosonic fields that reproduces 
the 3d reduced Lagrangian of the corresponding magical supergravity. 
This provides a unified treatment of all the magical supergravity theories in 
finding explicit relations between the 3d dimensionally reduced Lagrangians 
and particular coset nonlinear sigma models.
We also verify that the commutation relations of $E_{6(+2)}$, 
the quasi-conformal group for $\AA=\CC$, indeed satisfy this property, 
allowing the algebraic interpretation of the structure constants and 
scalar field functions as was done in the $F_{4(+4)}$ 
magical supergravity.

\end{abstract}

\preprint{KEK-TH 2038 }
\date{Feb.15, 2018}

\maketitle

\section{Introduction}
One of the remarkable discoveries in the history of supergravity theories 
is that of the existence of five-dimensional exceptional supergravities 
associated with the Freudenthal-Tits magic square \cite{GSTPhysLett,GSTNuclPhys}. 
They are a special class of $D=5$ $N=2$ Einstein-Maxwell supergravity 
(having 8 supercharges) that exist in addition to 
the infinite series of the non-Jordan family of $D=5$ $N=2$ supergravity. 
There are four such theories associated with the four division algebras, 
whose scalar manifolds 
arising in the reductions to $D=4$ and $3$ are 
coset manifolds of various (real forms of) exceptional and 
non-exceptional Lie groups, which, surprisingly enough, 
coincide with every entry of the table of the magic 
square that shows the symmetries of Jordan algebras (Table \ref{MagicSquare}). 
Since the discovery many works have been done 
 on these mysterious supergravity theories.  An incomplete list 
 includes \cite{GSTgauging,GSTmore,
 CFG,
 dWvP1,
 dWvP2,
 GZ, 
 GKN2,
 GKN,
 GP1,
 GP2,
 GNPW1,
 GP3,
 GNPW2,
 GNPW3,
 GunaydinPavlyk2,
 FerraraMarraniZumino,
 Karndumri1,
 Karndumri2,
 KanMizoguchi,
 Borsten:2017uoi}. (See also \cite{GunaydinReview} for 
a review.)
 
More recently, a precise identification was made in \cite{KanMizoguchi} 
between the bosonic fields of the three-dimensional reduced  
smallest magical supergravity and  
the parameter functions of  the coset space 
$F_{4(+4)}/(USp(6)\times SU(2))$, thereby
the $FFA$ couplings $C_{IJK}$ of the magical supergravity 
were shown to be identifiable as particular structure constants of 
the quasi-conformal algebra of the relevant Jordan algebra. 
It was also clarified there that 
the scalar fields $\stackrel{\circ}a\!{}^{IJ}$ and $\stackrel{\circ}a\!{}_{IJ}$
are nothing but the metric of the reduced dimensions 
in a particular representation.  Since 
the form of the dimensionally reduced Lagrangian is common to 
all magical supergravities, with the only differences being 
the range of the values of the indices of the vector and scalar fields,
it was conjectured in \cite{KanMizoguchi} that such a Lie algebraic 
characterization of the coupling constants or the scalar fields 
also applies to other magical supergravities, not only to the smallest 
${\rm J}^{\mathbb{R}}_3$ magical supergravity.

In this paper, we first show that 
if the commutation relations among the generators belonging to the 
respective irreducible components in the super-Ehlers decomposition 
are in a particular form, then 
one can always find a parameterization of the group element in terms of 
various 3d bosonic fields that reproduces 
the 3d reduced Lagrangian of the corresponding magical supergravity. 
This provides a unified treatment of all the magical supergravity theories in 
finding explicit relations between the 3d dimensionally reduced Lagrangians 
and particular coset nonlinear sigma models.
This is done in section 2.

We then verify that the commutation relations of $E_{6(+2)}$, 
the quasi-conformal group for $\AA=\CC$, allows a decomposition whose 
generators indeed satisfy this property, which immediately 
proves that the 3d reduced $\AA=\CC$ magical supergravity 
consists of an $E_{6(+2)}/(SU(6)\times SU(2))$ nonlinear sigma model 
coupled to supergravity. This is done in section 3.

We should mention that $F_{4(+4)}$, 
the quasi-conformal group for $\AA=\RR$, has also been shown 
to have such a property \cite{KanMizoguchi}. We conjecture 
that the remaining quasi-conformal groups in the list
($E_{7(-5)}$ (for $\AA=\HH$) and $E_{8(-24)}$ (for $\AA=\OO$) )
also possess such a special algebraic structure.

%

%


%
%
   \begin{table}[htp]
  \centering\caption{The magic square \cite{GSTPhysLett}}
   \begin{tabular}{llllll} \hline
    ~~~~~d               & J                &                    &                    &                    & ~~~~~~~$\mathbb{A}$~~~~~ \\ \cline{2-5}
                    & ${\rm J}^{\mathbb{R}}_3$ & ${\rm J}^{\mathbb{C}}_3$ & ${\rm J}^{\mathbb{H}}_3$ & ${\rm J}^{\mathbb{O}}_3$ & ~~~~~~~ \\ \hline
    ~~~~~5 (compact)     & $SO(3)$            & $SU(3)$           & $USp(6)$           & $F_4$              & ~~~~~~~$\mathbb{R}$~~~~~ \\
    ~~~~~5 (non-compact)~~~~~ & $SL(3,\mathbb{R})$ & $SL(3,\mathbb{C})$ & $SU^*(6)$          & $E_{6(-26)}$       & ~~~~~~~$\mathbb{C}$~~~~~ \\
    ~~~~~4 (non-compact)~~~~~ & $Sp(6,\mathbb{R})$ & $SU(3,3)$          & $SO^*(12)$         & $E_{7(-25)}$       & ~~~~~~~$\mathbb{H}$~~~~~ \\
    ~~~~~3 (non-compact)~~~~~ & $F_{4(+4)}$          & $E_{6(+2)}$        & $E_{7(-5)}$        & $E_{8(-24)}$       & ~~~~~~~$\mathbb{O}$~~~~~ \\ \hline
   \end{tabular}
   \label{MagicSquare}
   \end{table}
\section{Explicit construction of 3d nonlinear sigma models using the super-Ehlers decomposition}

The bosonic Lagrangian of a magical supergravity associated 
with the division algebra $\AA$ is given by
\beqa
\label{67}
\mathcal{L}&=&
\frac{1}{2}E^{(5)}R^{(5)}
-\frac{1}{4}E^{(5)}{\stackrel{\circ}a\!{}}_{IJ}F^I_{MN}F^{JMN}
-\frac{1}{2}E^{(5)}s_{xy}(\partial_M\phi^x)(\partial^M\phi^y)\n
&&
+\frac{1}{6\sqrt{6}}C_{IJK}\epsilon^{MNPQR}F^I_{MN}F^J_{PQ}A^K_R.
\eeqa
$E^{(5)}$ is the determinant of the f\"{u}nfbein,  
and $R^{(5)}$ is the scalar curvature in $D=5$. 
$F^I_{MN}$ is the $I$th Maxwell field strength $2\partial_{[\mu}A_{\nu]}^I$,
where $I,J,\dots=1,2,\dots,n_{\AAsub}+1$ with 
\beqa
n_{\AAsub}=3(1+\mbox{dim}\mathbb{A})-1. 
\eeqa
${\stackrel{\circ}a\!{}}_{IJ}$ and $s_{xy}$ are functions of $n_{\AAsub}$
scalar fields $\phi^x$ 
satisfying the relations ${\stackrel{\circ}a\!{}}_{IJ}={\stackrel{\circ}a\!{}}_{JI}$ 
and $s_{xy}=s_{yx}$.

Following a standard procedure of dimensional reduction and field dualization,  
one finds a 3d dimensionally reduced dualized Lagrangian for all magical supergravities as \cite{KanMizoguchi}
\begin{align}
\tilde{\mathcal{L}}~
=~&\frac{1}{2}ER
+\frac{1}{8}E\partial_\mu g^{mn}\partial^\mu g_{mn}
-\frac{1}{2}E e^{-2}\partial_\mu e\partial^\mu e
-\frac{1}{2}Es_{xy}(\partial_\mu\phi^x)(\partial^\mu\phi^y)
-\frac{1}{2}E{\stackrel{\circ}a\!{}}_{IJ}g^{mn}\partial_\mu A^I_m\partial^\mu A^J_n\nonumber\\
&-2Ee^{-2}{\stackrel{\circ}a\!{}}{}^{II'}
\left(\partial_\mu\varphi_I-\frac{1}{\sqrt{6}}C_{IJK}\epsilon^{mn}\partial_\mu A_m^JA_n^K\right)
\left(\partial^\mu\varphi_{I'}-\frac{1}{\sqrt{6}}C_{I'J'K'}\epsilon^{m'n'}\partial^\mu A_{m'}^{J'}A_{n'}^{K'}\right)\nonumber\\
&-Ee^{-2}g^{mn}
\left(
\partial_\mu\psi_m+\partial_\mu A^I_m\varphi_I-A^I_m\partial_\mu\varphi_I
+\frac{2}{3\sqrt{6}}C_{IJK}\epsilon^{pq}\partial_\mu A^I_pA^J_qA_m^K
\right)
\nonumber\\
&\qquad\qquad\times
\left(\partial^\mu\psi_n+\partial^\mu A^{I'}_n\varphi_{I'}-A^{I'}_n\partial^\mu\varphi_{I'}
+\frac{2}{3\sqrt{6}}C_{I'J'K'}\epsilon^{p'q'}\partial^\mu A^{I'}_{p'}A^{J'}_{q'}A^{K'}_n
\right),
\label{reducedLagrangian}
\end{align}
where $\mu,\nu,\ldots$ are the 3d spacetime indices and $m,n,m',n',\ldots$ are 
the reduced two-dimensional indices.
Note that this form is common to all the four magical supergravities; the only difference 
is the ranges the indices $x,y,\ldots$ and $I,J,\ldots$ run over.

Every magical supergravity contains in its 5d Lagrangian a nonlinear sigma model  
associated with the coset 
$\frac{\mbox{Str}_0({\rm J}_3^{\AAsub})}{\mbox{Aut}({\rm J}_3^{\AAsub})}$ 
\cite{GSTPhysLett,GSTNuclPhys}. When it is dimensionally reduced to 
three dimensions, the scalar coset is enlarged to 
$\frac{\mbox{qConf}({\rm J}_3^{\AAsub})}{\widetilde{\mbox{M\"{o}}}({\rm J}_3^{\AAsub})
\times SU(2)}$,  in which all the non-gravity bosonic degrees of freedom  
are contained.
To show explicitly how the various terms arising through the dimensional 
reduction are gathered to form a single coset, it is convenient to decompose 
the Lie algebra of 
the quasi-conformal group $\mbox{qConf}({\rm J}_3^{\AAsub})$, which 
is the numerator group of the 3d coset, in terms of representations of 
the Lie algebra of the subgroup $SL(3,\RR)\times \mbox{Str}_0({\rm J}_3^{\AAsub})$,
the latter of which is the numerator group of the 5d coset. The decomposition 
is always in the same form for all magical supergravities\cite{FerraraMarraniZumino}:
\beqa
\mbox{qConf}({\rm J}_3^{\AAsub})&\supset& SL(3,\RR)\times \mbox{Str}_0({\rm J}_3^{\AAsub}),\n
\mbox{\bf adj}(\mbox{qConf}({\rm J}_3^{\AAsub}))
&=&({\bf 8},{\bf 1})\oplus
({\bf 3},{\bf \overline{n_{\AAsub}+1}})
\oplus
({\bf \bar 3},{\bf n_{\AAsub}+1})
\oplus({\bf 1},\mbox{\bf adj}(\mbox{Str}_0({\rm J}_3^{\AAsub}))).
\label{qConfdecomposition}
\eeqa

We will show that if 
the generators of the Lie algebra of 
the quasi-conformal group $\mbox{qConf}({\rm J}_3^{\AAsub})$ 
take a particular form (\ref{commutationrelations}) 
as assumed below, then one can always 
reproduce the Lagrangian (\ref{reducedLagrangian})
as a coset nonlinear sigma model coupled to gravity.

Let the generators of the Lie algebra of 
the quasi-conformal group $\mbox{qConf}({\rm J}_3^{\AAsub})$ 
satisfy 
 \begin{eqnarray}
{[}\hat{E}^{i}_{~j},~\hat{E}^{k}_{~l}{]} 
&=& \delta^{k}_{j}\hat{E}^{i}_{~l}-\delta^{i}_{l}\hat{E}^{k}_{~j},  \n
{[}\hat{E}^{i}_{~j},{E}^{*k}_{I}{]} 
&=&\delta_j^k E^{*i}_{I},\n
{[}\hat{E}^{i}_{~j},~{E}_{k}^{I}{]} 
&=&-\delta_k^i E_{j}^{I},\n
{[}T_{\tilde i},~T_{\tilde j}{]} 
&=&f_{{\tilde i}{\tilde j}}^{~~\tilde k}T_{\tilde k},\n
{[}T_{\tilde i},{E}^{*k}_{I}{]} 
&=&{\bar t}_{{\tilde i}I}^{~~J}{E}^{*k}_{J} ,\n
{[}T_{\tilde i},~{E}_{k}^{I}{]} 
&=&t_{{\tilde i}~J}^{~I}{E}_{k}^{J} ,\n
{[}{E}_{i}^{I},~{E}_{j}^{J}{]} 
&=&C^{IJK}\epsilon_{ijk}{E}^{*k}_{K} ,\n
{[}{E}^{*i}_{I},~{E}^{*j}_{J}{]} 
&=&-C_{IJK}\epsilon^{ijk}{E}_{k}^{K},\n
{[}{E}_{i}^{I},~{E}^{*j}_{J}{]} 
&=&-2\delta^I_J\hat{E}^j_{~i} +\delta^i_j D^{I~{\tilde i}}_{~J} T_{\tilde i},\n
\mbox{otherwise}&=&0.
\label{commutationrelations}
 \end{eqnarray}
$\epsilon^{ijk}$ and $\epsilon_{ijk}$ are completely antisymmetric tensors with 
$\epsilon_{123}=\epsilon^{123}=1$.

$\hat{E}^{i}_{~j}$ ($i,j=1,\ldots,3$) with 
$\hat{E}^{1}_{~1}+\hat{E}^{2}_{~2}+\hat{E}^{3}_{~3}=0$ 
are the generators of $SL(3,\RR)$,  
the first irreducible component  $({\bf 8},{\bf 1})$ of (\ref{qConfdecomposition}). 
They are defined modulo 
$\hat{E}^{1}_{~1}+\hat{E}^{2}_{~2}+\hat{E}^{3}_{~3}$, 
that is, $\hat{E}^{i}_{~j}$ is an element of a quotient space 
of $GL(3,\RR)$ divided by the center generated by the
overall $U(1)$ generator. 
$T_{\tilde i}$ (${\tilde i}=1,\ldots,\mbox{dim}\mbox{Str}_0({\rm J}_3^{\AAsub})$) 
are the generators of $\mbox{Str}_0({\rm J}_3^{\AAsub})$ of the respective 
magical supergravity, which is the last irreducible component of 
(\ref{qConfdecomposition}).
Finally, 
${E}^{*j}_{J}$ and ${E}_{i}^{I}$ ($i,j=1,\ldots,3;~ I,J=1,\ldots,n_{\AAsub}+1$) are the 
generators of  
$({\bf 3},{\bf \overline{n_{\AAsub}+1}})$
and $({\bf \bar 3},{\bf n_{\AAsub}+1})$, respectively.

$f_{{\tilde i}{\tilde j}}^{~~\tilde k}$, 
${\bar t}_{{\tilde i}I}^{~~J}$,
$t_{{\tilde i}~J}^{~I}$,
$C_{IJK}$,$C^{IJK}$ and 
$D^{I~{\tilde i}}_{~J}$ 
are {\em real} structure constants; they are not all independent 
but are constrained by the Jacobi identities. The full set of constraints are
\beqa
C^{IJK}&=&C^{JIK}=C^{JKI}=C^{KIJ},\\
C_{IJK}&=&C_{JIK}=C_{JKI}=C_{KIJ},\\
{\bar t}_{{\tilde i}I}^{~~J}&=&
-t_{{\tilde i}~I}^{~J},
\label{tiIJ=-tiIJ}\\
t_{{\tilde i}~K}^{~I}C^{KJL}+
t_{{\tilde i}~K}^{~J}C^{KLI}+
t_{{\tilde i}~K}^{~L}C^{KIJ}&=&0,\\
t_{{\tilde i}~I}^{~K}C_{KJL}+
t_{{\tilde i}~J}^{~K}C_{KLI}+
t_{{\tilde i}~L}^{~K}C_{KIJ}&=&0,\\
D^{I~{\tilde i}}_{~K}t_{{\tilde i}~M}^{~J}-
D^{J~{\tilde i}}_{~K}t_{{\tilde i}~M}^{~I}&=&
2(\delta^{J}_K \delta^{I}_M-
\delta^{I}_K \delta^{J}_M),\\
D^{I~{\tilde i}}_{~K}t_{{\tilde i}~M}^{~J}+
D^{J~{\tilde i}}_{~K}t_{{\tilde i}~M}^{~I}&=&
2(\delta^{J}_K \delta^{I}_M+
\delta^{I}_K \delta^{J}_M-
C^{IJL} C_{LKM}),\\
t_{{\tilde i}~I}^{~K} D^{J~{\tilde j}}_{~K}-
t_{{\tilde i}~K}^{~J} D^{K~{\tilde j}}_{~I}
&=& D^{J~{\tilde k}}_{~I} f_{{\tilde k}{\tilde i}}^{~~\tilde j},\\
{[} t_{\tilde i},~t_{\tilde j}{]}^I_{~J}&=&-f_{{\tilde i}{\tilde j}}^{~~\tilde k}t_{{\tilde k}~J}^{~I}.
\eeqa
Note that the symmetricity of $C_{IJK}$ or $C^{IJK}$ is required 
from the Jacobi identities of the above algebra. Of course, 
this property is one of the virtues of magical supergravity theories.

$\mbox{Str}_0({\rm J}_3^{\AAsub})$ and 
$\mbox{qConf}({\rm J}_3^{\AAsub})$ are symmetric spaces
for all ${\AA}$. 
The symmetric space involution $\tau$ defined on the former coincides with 
that on the latter if regarded as the action on its subgroup 
$\mbox{Str}_0({\rm J}_3^{\AAsub})$. 
Using this fact, one can show that 
\beqa
{\bf H}&=&(\oplus_{i,j=1,2,3} \RR (\hat E^i_{~j}-\hat E^j_{~i}))
\n
&&
\oplus
(\oplus_{i=1,2,3;I=1,\ldots,n_{\AAsub}+1} \RR (E_i^{I}-E_I^{*i}))
\n
&&
\oplus
{\bf H}_{{\rm Str}_0({\rm J}_3^{\AAsub})}
\eeqa
and
\beqa
{\bf K}&=&(\oplus_{i,j=1,2,3} \RR (\hat E^i_{~j}+\hat E^j_{~i}))
\n
&&
\oplus
(\oplus_{i=1,2,3;I=1,\ldots,n_{\AAsub}+1} \RR (E_i^{I}+E_I^{*i}))\n
&&
\oplus
{\bf K}_{{\rm Str}_0({\rm J}_3^{\AAsub})}
\label{HandK}
\eeqa 
satisfy
\beqa
{[}{\bf H},~{\bf H}{]}&\subset&{\bf H},\n
{[}{\bf K},~{\bf K}{]}&\subset&{\bf H},\n
{[}{\bf H},~{\bf K}{]}&\subset&{\bf K},
\eeqa 
so that one can define the symmetric space involution $\tau$ as 
\beqa
\tau({\bf H})=+{\bf H},~~~\tau({\bf K})=-{\bf K}.
\eeqa

Now the construction of the coset sigma model is 
straightforward. 
Defining
\beqa
{\cal V}&=&{\cal V}_-{\cal V}_+,\\
{\cal V}_+&=&{\cal V}_+^{(grav)}{\cal V}_+^{(scalar)},\\
{\cal V}_+^{(grav)}&=&
\exp\left(
\log e_{\dot 1}^{~1}\hat E^1_{~1}
+\log e_{\dot 2}^{~2}\hat E^2_{~2}
+\log e\;\hat E^3_{~3}
\right)\n
&&\cdot\exp\left(
-e_1^{~\dot 2} e_{\dot 2}^{~2} \hat E^1_{~2}\right)
\exp\left(
\psi_1\hat E^1_{~3}+\psi_2\hat E^2_{~3}
\right),
\\
 {\cal V}_+^{(scalar)}&=&\exp\left((\log{\bf\tilde  e}^{-1})^{\tilde i}
 T_{\tilde i}\right)~~~~({\tilde i}=1,\ldots,\mbox{dim}\mbox{Str}_0({\rm J}_3^{\AAsub})) ,
\eeqa
\beqa
{\cal V}_-&=&\exp\left(
A_{m}^I E_I^{*m}
+\varphi_I E^I_3\right)
~~~(m=1,2; ~ I=1,\ldots,n_{\AAsub}+1),
\eeqa
the Maurer-Cartan 1-form is computed as 
\beqa
\partial_\mu {\cal V}{\cal V}^{-1}&=&
\partial_\mu {\cal V}_+{\cal V}_+^{-1}
+ {\cal V}_+(\partial_\mu {\cal V}_-{\cal V}_-^{-1}){\cal V}_+^{-1},\\
\partial_\mu {\cal V}_+{\cal V}_+^{-1}&=&
(e_{\dot 1}^{~1})^{-1}\partial_\mu e_{\dot 1}^{~1} \hat E^1_{~1}+
(e_{\dot 2}^{~2})^{-1}\partial_\mu e_{\dot 2}^{~2} \hat E^2_{~2}+
e^{-1}\partial_\mu e \hat E^3_{~3}\n
&&-e_{\dot 1}^{~1}(e_{\dot 2}^{~2})^{-1} \partial_\mu B\hat E^1_{~2}
+e^{-1}\left(
e_{\dot 1}^{~1}(\partial_\mu \psi_1-B\partial_\mu \psi_2)\hat E^1_{~3}
+e_{\dot 2}^{~2}\partial_\mu \psi_2 \hat E^2_{~3}
\right)\n
&&+(\partial_\mu {\bf \tilde e}^{-1}\cdot{\bf\tilde  e})^{\tilde i}
T_{\tilde i}
\label{dV+V+-1},
\eeqa
where
\beqa
e_{m}^{~\dot m}&=&\left(
\begin{array}{cc} e_{1}^{~\dot 1}&e_{1}^{~\dot 2}\\
0&e_{2}^{~\dot 2}
\end{array}
\right),
\eeqa
\beqa
({\bf e}^{-1})_a^{~i}&=&\left(
\begin{array}{cc}
e_{\dot{m}}^{~m}&e_{\dot{m}}^{~m}\psi_{m}\\
0&e
\end{array}
\right),
\eeqa
\beqa
e&=&{\rm det}e_{m}^{~\dot{m}}
=(e_{\dot 1}^{~1}e_{\dot 2}^{~2})^{-1}
\eeqa
for the ${\cal V}_+$ term, whereas 
\beqa
\partial_\mu {\cal V}_-{\cal V}_-^{-1}
&=&
\partial_\mu A_{m}^I E_I^{*m}+\left(
\partial_\mu\varphi_I-\frac12 C_{JKI}\epsilon^{mn}A_{m}^J\partial_\mu A_{n}^K
\right)E^K_3\n
&&+
\left(
A_{m}^I\partial_\mu\varphi_I- \partial_\mu A_{m}^I ~\varphi_I
-\frac{1}3 C_{JKI} \epsilon^{np}A_{m}^I A_{n}^J\partial_\mu A_{p}^K 
\right)\hat E^{m}_{~~3}
\eeqa 
for the ${\cal V}_-$ term. Here $m,n,p$ are the two-dimensional 
curved indices taking 1 or 2, while $\dot{m}$ is the two-dimensional 
flat index taking $\dot{1}$ or $\dot{2}$.

We can compute the ${\cal V}_+^{(grav)}$ conjugations as 
\beqa
{\cal V}_+^{(grav)}
E_I^{*j}
{\cal V}_+^{(grav)-1}
&=&
E_I^{*a}({\bf e}^{-1})_a^{~j},
\n
{\cal V}_+^{(grav)}
E^I_{j}
{\cal V}_+^{(grav)-1}
&=&
({\bf e})_j^{~a}E^I_{a},
\n
{\cal V}_+^{(scalar)}
E_I^{*j}
{\cal V}_+^{(scalar)-1}
&=&
\stackrel{\circ}f\!{}_I^{~A}
E_A^{*j},
\n
{\cal V}_+^{(scalar)}
E^I_{j}
{\cal V}_+^{(scalar)-1}
&=&
\stackrel{\circ}f\!{}^I_{~A}E^A_{j},
\eeqa
where 
\beqa
\stackrel{\circ}f\!{}_I^{~A}&=&(\exp((\log \tilde {\bf e}^{-1})^{\tilde i}
\bar t_{\tilde i}))_I^{~A},\\
\stackrel{\circ}f\!{}^I_{~A}&=&(\exp((\log \tilde {\bf e}^{-1})^{\tilde i}
t_{\tilde i}))^I_{~A}.
\eeqa
Note that, due to the relation 
${\bar t}_{{\tilde i}I}^{~~J}=-t_{{\tilde i}~I}^{~J}$ 
(\ref{tiIJ=-tiIJ}),
$\stackrel{\circ}f\!{}_I^{~A}$ is the transpose of 
the inverse of $\stackrel{\circ}f\!{}^I_{~A}$.
Thus we find 
\beqa
 {\cal V}_+(\partial_\mu {\cal V}_-{\cal V}_-^{-1}){\cal V}_+^{-1}
&=&
e_{\dot{m}}^{~m}\stackrel{\circ}f\!{}_I^{~A}\partial_\mu A_{m}^I E_A^{*\dot{m}}\n
&&+e^{-1}\stackrel{\circ}f\!{}^I_{~A}\left(
\partial_\mu\varphi_I-\frac12 C_{JKI}\epsilon^{mn}A_{m}^J\partial_\mu A_{n}^K
\right)E^A_3\n
&&+e^{-1}e_{\dot{m}}^{~m}
\left(
A_{m}^I\partial_\mu\varphi_I- \partial_\mu A_{m}^I ~\varphi_I
-\frac{1}3 C_{JKI} \epsilon^{np}A_{m}^I A_{n}^J\partial_\mu A_{p}^K 
\right)\hat E^{\dot{m}}_{~~3}.\n
\label{V+dV-V--1V--1}
\eeqa 
%
Defining
$
{\cal M}\equiv\tau({\cal V}^{-1}){\cal V}
$
as usual, 
we have
\beqa
-\frac14 E^{(3)}\mbox{Tr}\partial_\mu{\cal  M}^{-1} \partial^\mu{\cal M}
&=&E^{(3)}\mbox{Tr}\left(\frac12\left(
\partial_\mu{\cal V}{\cal V}^{-1}-\tau\left(\partial_\mu{\cal V}{\cal V}^{-1}\right)
\right)\right)^2,
\label{TrdMdM-1}
\eeqa
thereby the {\bf H} pieces of $\partial_\mu {\cal V}{\cal V}^{-1}$ are projected out. 
This amounts to the replacements 
of the generators in $\partial_\mu{\cal V}{\cal V}^{-1}$ as
\beqa
\hat E^i_{~j}&\longrightarrow&\frac12(\hat E^i_{~j}+\hat E^j_{~i}),\n
E^I_i&\longrightarrow&\frac12(E^I_i+E_I^{*i}),\n
E_I^{*i}&\longrightarrow&\frac12(E^I_i+E_I^{*i}),\n
T_{\tilde i}&\longrightarrow&\frac12(T_{\tilde i}-\tau(T_{\tilde i})).
\eeqa
Using the traces with the normalizations 
\beqa
\frac1{2h}
{\rm Tr}{\hat E}^a_{~b}{\hat E}^c_{~d}&=&\delta^c_b \delta^a_d~~~(a,b,c,d=1,2,3),
\n
\frac1{2h}{\rm Tr}E^A_{a}E_B^{*b}&=&
2\delta^b_a \delta^A_B~~~(a,b=1,2,3;~~A,B=1,\ldots,n_{\AAsub}+1) ,
\n
\frac1{2h}
{\rm Tr}(T_{\tilde i}-\tau(T_{\tilde i}))(T_{\tilde j}-\tau(T_{\tilde j})
&=&\gamma_{{\tilde i}{\tilde j}}
~~~({\tilde i},{\tilde j}=1,\ldots,\mbox{dim}\mbox{Str}_0({\rm J}_3^{\AAsub})) ),
\n
\mbox{otherwise}&=&0,
\eeqa
where $h$ is the dual Coxeter number of $\mbox{qConf}({\rm J}_3^{\AAsub})$,
we obtain the final result
\beqa
\frac1{8h}{\rm Tr}\partial_\mu{\cal M}^{-1}
\partial^\mu {\cal M}
&=&\frac14 \partial_\mu g^{mn} \partial_\mu g_{mn} 
-e^{-2}\partial_\mu e \partial^\mu e
- g^{mn}\!\stackrel{\circ}a\!{}_{IJ} \partial_\mu A_m^I
 \partial^\mu A_n^J\n
 &&+\frac12  
 \gamma_{\tilde i\tilde j} (\partial_\mu {\bf \tilde e}^{-1}\cdot{\bf\tilde  e})^{\tilde i}
 (\partial^\mu {\bf \tilde e}^{-1}\cdot{\bf\tilde  e})^{\tilde j}
\n
&&- e^{-2}
\stackrel{\circ}a\!{}^{IJ}
\left(
\partial_\mu\varphi_I
-\frac12 C_{KLI}\epsilon^{mn}A^K_m \partial_\mu A^L_n
\right)\n
&&
~~~~~~~~~~~~\cdot
\left(
\partial^\mu\varphi_J
-\frac12 C_{K'L'J}\epsilon^{m'n'}A^K_{m'} \partial_\mu A^L_{n'}
\right)
\n
&&-\frac12 e^{-2} g^{mn} 
\left(
\partial_\mu\psi_m -\varphi_I\partial_\mu A^I_m+\partial_\mu\varphi\; A^I_m
-\frac13 C_{KLI}\epsilon^{np} A^K_n \partial_\mu A^L_p \; A^I_m
\right)\n
&&~~~~~~~\cdot \left(
\partial^\mu\psi_n -\varphi_I\partial_\mu A^J_n+\partial_\mu\varphi\; A^J_n
-\frac13 C_{K'L'J}\epsilon^{n'p'} A^{K'}_{n'} \partial_\mu A^{L'}_{p'}A^J_n
\right).\n
&&
\label{finalsigmamodel}
\eeqa
The second line is the $\frac{\mbox{Str}_0({\rm J}_3^{\AAsub})}{\mbox{Aut}({\rm J}_3^{\AAsub})}$ 
sigma model contained in the original $D=5$ magical supergravity Lagrangian.
The result (\ref{finalsigmamodel}) coincides with the sigma-model 
part of (\ref{reducedLagrangian}) after appropriate rescalings of $\varphi_I, \psi_m$ 
and $C_{IJK}$.\footnote{There was a typo in the formula of rescalings 
(64) in \cite{KanMizoguchi}.}

\section{$E_{6(+2)}$ algebra}
$E_{6(+2)}$ is one of the real forms of the exceptional Lie algebra $E_{6}$. 
This is different from the more familiar split real form $E_{6(+6)}$
encountered as a U-duality group in type II string theory, 
which is conveniently realized as a Lie subalgebra of $E_{8(+8)}$ 
by using the generators 
in Freudenthal's realization 
\cite{MizoguchiE10,MizoguchiSchroder,MizoguchiOhta}. 
Thus we first determine the generators of $E_{6(+6)}$ in $E_{8(+8)}$, 
take its complexification $E_{6(+6)}\otimes  \CC$, and then 
we identify the generators of the real Lie algebra $E_{6(+2)}$.

The  $E_{8(+8)}$ generators are  (The numbers in the parentheses
are the total numbers of the respective generators.)%
\beqa
\begin{array}{lll}
E^I_{~J} & (I,J=1,\ldots,9; ~~I\neq J) & \mbox{(72),}\\
E^{IJK} &(I,J,K=1,\ldots,9)&\mbox{(84),}\\
E^*_{IJK} &(I,J,K=1,\ldots,9)&\mbox{(84),}\\
h_I &(I=1,\ldots,8)~~(=E^I_{~I}-E^J_{~J})&\mbox{(8),}
\end{array}
\eeqa
which are assumed to satisfy the commutation relations
\footnote{ 
The sign factor of  ${[}E^*_{IJK},~~E^*_{LMN}{]}$ is different 
from that in ref.\cite{MizoguchiSchroder} because the 
metric used there was the 
``mostly negative" one.
}
\beqa
\begin{array}{lcl}
{[}E^I_{~J},~~E^K_{~L}{]}&=&\delta^{K}_{J} E^I_{~L} -\delta^{I}_{L} E^K_{~J},\\
{[}E^I_{~J},~~E^{KLM}{]}&=&3\delta^{[M}_{J}E^{KL]I},\\
{[}E^I_{~J},~~E^*_{KLM}{]}&=&-3\delta^{I}_{[M}E^*_{KL]J},\\
{[}E^{IJK},~~E^{LMN}{]}&=&-\frac1{3!}\sum_{P,Q,R=1}^9 \epsilon^{IJKLMNPQR}E^*_{PQR},\\
{[}E^*_{IJK},~~E^*_{LMN}{]}&=&+\frac1{3!}\sum_{P,Q,R=1}^9 \epsilon_{IJKLMNPQR}E^{PQR},\\
{[}E^{IJK},~~E^*_{LMN}{]}&=&6\delta^J_{[M}\delta_N^K E^I_{~L]}~~~(\mbox{if $I\neq L,M,N$}),\\
{[}E^{IJK},~~E^*_{IJK}{]}&=&h_{IJK},
\end{array}
\eeqa
where
\beqa
h_{IJK}&\equiv&E^I_{~I}+E^J_{~J}+E^K_{~K}-\frac13\sum_{L=1}^9 E^L_{~L}.
\eeqa
Among them, $E_{6(+6)}$ is generated by the following 78 generators: 
\beqa
\begin{array}{ll}
E^{\hat{i}}_{~\hat{j}}(1\leq \hat{i} \neq \hat{j} \leq 6), E^1_{~1}-E^2_{~2}, E^2_{~2}-E^3_{~3}, E^3_{~3}-E^4_{~4},  
E^4_{~4}-E^5_{~5}, E^5_{~5}-E^6_{~6}
& \mbox{(35),}\\
E^{\hat{i}\hat{j}\hat{k}}(1\leq \hat{i} <\hat{j}<\hat{k} \leq 6)&\mbox{(20),}\\
E^*_{\hat{i}\hat{j}\hat{k}}(1\leq \hat{i} <\hat{j}<\hat{k} \leq 6)&\mbox{(20),}\\
E^{789},  E^*_{789},  h_{789}
& \mbox{(3).}
\end{array}
\eeqa
The first line is the generators of the $SL(6,\RR)$ subalgebra, 
whereas the bottom line is the ones of the $SL(2,\RR)$ subalgebra.
They form a real Lie algebra with all real structure constants, whose 
complexification is the complex Lie algebra $E_6$. 
Among the complex generators thus obtained, 
we can find another set of generators forming a basis of a
 different real form of $E_6$ as follows:

\begin{enumerate}
  \item The generators 
  $E^{123},~E^{456},~E^{789},~E^{*}_{123},~E^{*}_{456},
	~E^{*}_{789},~h_{123}$  and  $h_{456}$ 
  form an $SL(3,\RR)$ algebra. We identify them 
  as the $\hat E^i_{~j}$ generators in the previous section 
  as  
  \beqa
  && \hat{E}^{1}_{~2}=E^*_{123},~\hat{E}^{2}_{~3}=E^*_{456},~\hat{E}^{1}_{~3}=E^{789},~\n
  && \hat{E}^{2}_{~1}=E^{123},~\hat{E}^{3}_{~2}=E^{456},~\hat{E}^{3}_{~1}=E^*_{789},~\n
  && \hat{E}^{1}_{~1}=-\frac13(2 h_{123}+h_{456}),
  ~\hat{E}^{2}_{~2}=\frac13(h_{123}-h_{456}).
 \label{Ehatgenerators}
  \eeqa
  (Note that we have defined the $\hat E^i_{~j}$ generators modulo 
  $\hat{E}^{1}_{~1}+\hat{E}^{2}_{~2}+\hat{E}^{3}_{~3}$, so for instance 
  $\hat{E}^{1}_{~1}$ is equal to 
  $\frac13(2\hat{E}^{1}_{~1}-\hat{E}^{2}_{~2}-\hat{E}^{3}_{~3})$.)
  \item Besides, there are two commuting $SL(3,\RR)$'s in the 
  $SL(6,\RR)$ subalgebra of $E_{6(+6)}$, and by complexification 
  one can construct two commuting sets of Gell-Mann matrices generating 
  $SU(3)\oplus SU(3)$. Let these generators be $\lambda_r$ and $\tilde\lambda_r$ 
  $(r=1,\ldots,8)$, then they can be taken to be
 \beqa
&& \lambda_1=E^1_{~2}+E^2_{~1},~
 \lambda_2=i(-E^1_{~2}+E^2_{~1}),~
 \lambda_3=E^1_{~1}-E^2_{~2},~
 \lambda_4=E^2_{~3}+E^3_{~2},~\n
&& \lambda_5=i(-E^2_{~3}+E^3_{~2}),~
 \lambda_6=E^1_{~3}+E^3_{~1},~
 \lambda_7=i(-E^1_{~3}+E^3_{~1}),~
 \lambda_8=E^2_{~2}-E^3_{~3},~~~\\
 && \tilde\lambda_1=E^4_{~5}+E^5_{~4},~
 \tilde\lambda_2=i(-E^4_{~5}+E^5_{~4}),~
 \tilde\lambda_3=E^4_{~4}-E^5_{~5},~
 \tilde\lambda_4=E^5_{~6}+E^6_{~5},~\n
&& \tilde\lambda_5=i(-E^5_{~6}+E^6_{~5}),~
 \tilde\lambda_6=E^4_{~6}+E^6_{~4},~
 \tilde\lambda_7=i(-E^4_{~6}+E^6_{~4}),~
 \tilde\lambda_8=E^5_{~5}-E^6_{~6},~~~
 \eeqa 
where $\lambda_8$ and $\tilde \lambda_8$ are not what should 
correspond to the original eighth Gell-Mann matrix, but are 
generators corresponding to 
$\left(
\begin{array}{ccc}
0&&\\
&1&\\
&&-1
\end{array}
\right)$ for respective $SU(3)$ algebras.
 Defining 
\beqa
\mu_r=i(\lambda_r + \tilde \lambda_r),~~~
\nu_r=-\lambda_r + \tilde \lambda_r~~~(r=1,\ldots,8),
\eeqa
then we see that these 16 generators form a basis of a {\em real} 
Lie algebra $SL(3,\CC)$. 

As the $T_{\tilde i}$ generators for 
$\mbox{Str}_0({\rm J}_3^{\CCsub})=SL(3,\CC)$, 
it is convenient to consider another set of 16 generators 
defined by taking the following real linear combinations of $\mu_r$ and $\nu_r$:

 \beqa
 \begin{array}{rclcl}
T_{1} ~ &=&\frac13\left( 2\nu_3 +  \nu_8\right) &=&
 ~ \frac{1}{3}(-2E^{1}_{~1}+E^{2}_{~2}+E^{3}_{~3}+2E^{4}_{~4}-E^{5}_{~5}-E^{6}_{~6}),\\
T_{2} ~ &=&\frac12\left( \mu_2 + \nu_1\right) &=&
 ~ -E^{2}_{~1}+E^{4}_{~5},\\
T_{3} ~ &=&\frac12\left( \mu_7 + \nu_6\right) &=&
 ~ -E^{3}_{~1}+E^{4}_{~6},\\
T_{4} ~ &=&\frac12\left( -\mu_2 + \nu_1\right) &=&
 ~ -E^{1}_{~2}+E^{5}_{~4},\\
T_{5} ~ &=&\frac13\left( -\nu_3 +  \nu_8\right) &=&
 ~ \frac{1}{3}(E^{1}_{~1}-2E^{2}_{~2}+E^{3}_{~3}-E^{4}_{~4}+2E^{5}_{~5}-E^{6}_{~6}),\\
T_{6} ~ &=&\frac12\left( \mu_5 + \nu_4\right) &=&
 ~ -E^{3}_{~2}+E^{5}_{~6},\\
T_{7} ~ &=&\frac12\left(-\mu_7 + \nu_6\right) &=&
 ~ -E^{1}_{~3}+E^{6}_{~4},\\
T_{8} ~ &=&\frac12\left( -\mu_5 + \nu_4\right) &=&
 ~ -E^{2}_{~3}+E^{6}_{~5},\\
T_{9}  ~ &=&\frac13\left( 2\mu_3 +  \mu_8\right) &=&
 ~ \frac{i}{3}(2E^{1}_{~1}-E^{2}_{~2}-E^{3}_{~3}+2E^{4}_{~4}-E^{5}_{~5}-E^{6}_{~6}),\\
T_{10} ~ &=&\frac12\left( \mu_1 - \nu_2\right) &=&
 ~ i(E^{2}_{~1}+E^{4}_{~5}),\\
T_{11} ~ &=&\frac12\left( \mu_6 - \nu_7\right) &=&
 ~ i(E^{3}_{~1}+E^{4}_{~6}),\\
T_{12} ~ &=&\frac12\left( \mu_1 + \nu_2\right) &=&
 ~ i(E^{1}_{~2}+E^{5}_{~4}),\\
T_{13} ~ &=&\frac13\left(-\mu_3 +  \mu_8\right) &=&
 ~ \frac{i}{3}(-E^{1}_{~1}+2E^{2}_{~2}-E^{3}_{~3}-E^{4}_{~4}+2E^{5}_{~5}-E^{6}_{~6}),\\
T_{14} ~ &=&\frac12\left(\mu_4 - \nu_5\right) &=&
 ~ i(E^{3}_{~2}+E^{5}_{~6}),\\
T_{15} ~ &=&\frac12\left(\mu_6 + \nu_7\right) &=&
 ~ i(E^{1}_{~3}+E^{6}_{~4}),\\
T_{16} ~ &=&\frac12\left(\mu_4 + \nu_5\right) &=&
 ~ i(E^{2}_{~3}+E^{6}_{~5}).
\end{array}
 \n
 \label{Tgenerators}
 \eeqa
\item
Finally, the generators transforming as 
$({\bf 3},{\bf \overline{n_{\AAsub}+1}})
\oplus
({\bf \bar 3},{\bf n_{\AAsub}+1})
$ with $n_{\AAsub}=8$ under 
$\mbox{Str}_0({\rm J}_3^{\CCsub})=SL(3,\CC)$ 
(\ref{qConfdecomposition}) 
are identified to be :

\par
$E^{*i}_{I}$ :
\beqa
 \begin{array}{lcllcllcl}
E^{*1}_{1} ~ &=& ~ \sqrt{2}E^{*}_{234} ,&
E^{*2}_{1} ~ &=& ~ \sqrt{2}E^{1}_{~4},&
E^{*3}_{1} ~ &=& ~ -\sqrt{2}E^{156},\n
E^{*1}_{2} ~ &=& ~ \sqrt{2}E^{*}_{315} ,&
E^{*2}_{2} ~ &=& ~ \sqrt{2}E^{2}_{~5},&
E^{*3}_{2} ~ &=& ~ -\sqrt{2}E^{264},\n
E^{*1}_{3} ~ &=& ~ \sqrt{2}E^{*}_{126} ,&
E^{*2}_{3} ~ &=& ~ \sqrt{2}E^{3}_{~6},&
E^{*3}_{3} ~ &=& ~ -\sqrt{2}E^{345},\n
E^{*1}_{4} ~ &=& ~ E^{*}_{316}+E^{*}_{125},&
E^{*2}_{4} ~ &=& ~ E^{2}_{~6}+E^{3}_{~5},&
E^{*3}_{4} ~ &=& ~ -(E^{364}+E^{245}),\n
E^{*1}_{5} ~ &=& ~ E^{*}_{124}+E^{*}_{236},&
E^{*2}_{5} ~ &=& ~ E^{3}_{~4}+E^{1}_{~6},&
E^{*3}_{5} ~ &=& ~ -(E^{145}+E^{356}),\n
E^{*1}_{6} ~ &=& ~ E^{*}_{235}+E^{*}_{314},&
E^{*2}_{6} ~ &=& ~ E^{1}_{~5}+E^{2}_{~4},&
E^{*3}_{6} ~ &=& ~ -(E^{256}+E^{164}),\n
E^{*1}_{7} ~ &=& ~ i(E^{*}_{316}-E^{*}_{125}),&
E^{*2}_{7} ~ &=& ~ i(E^{2}_{~6}-E^{3}_{~5}),&
E^{*3}_{7} ~ &=& ~ i(E^{364}-E^{245}),\n
E^{*1}_{8} ~ &=& ~ i(E^{*}_{124}-E^{*}_{236}),&
E^{*2}_{8} ~ &=& ~ i(E^{3}_{~4}-E^{1}_{~6}),&
E^{*3}_{8} ~ &=& ~ i(E^{145}-E^{356}),\n
E^{*1}_{9} ~ &=& ~ i(E^{*}_{235}-E^{*}_{314}),&
E^{*2}_{9} ~ &=& ~ i(E^{1}_{~5}-E^{2}_{~4}),&
E^{*3}_{9} ~ &=& ~ i(E^{256}-E^{164}).
 \end{array}\n
 \label{Estargenerators}
 \eeqa

$E^{I}_{i}$ :
\beqa
 \begin{array}{lcllcllcl}
E^{1}_{1} ~ &=& ~ \sqrt{2}E^{234},&
E^{1}_{2} ~ &=& ~ \sqrt{2}E^{4}_{~1},&
E^{1}_{3} ~ &=& ~ -\sqrt{2}E^{*}_{156},\n
E^{2}_{1} ~ &=& ~ \sqrt{2}E^{315},&
E^{2}_{2} ~ &=& ~ \sqrt{2}E^{5}_{~2},&
E^{2}_{3} ~ &=& ~ -\sqrt{2}E^{*}_{264},\n
E^{3}_{1} ~ &=& ~ \sqrt{2}E^{126},&
E^{3}_{2} ~ &=& ~ \sqrt{2}E^{6}_{~3},&
E^{3}_{3} ~ &=& ~ -\sqrt{2}E^{*}_{345},\n
E^{4}_{1} ~ &=& ~ E^{125}+E^{316},&
E^{4}_{2} ~ &=& ~ E^{5}_{~3}+E^{6}_{~2},&
E^{4}_{3} ~ &=& ~ -(E^{*}_{364}+E^{*}_{245}),\n
E^{5}_{1} ~ &=& ~ E^{236}+E^{124},&
E^{5}_{2} ~ &=& ~ E^{6}_{~1}+E^{4}_{~3},&
E^{5}_{3} ~ &=& ~ -(E^{*}_{145}+E^{*}_{356}),\n
E^{6}_{1} ~ &=& ~ E^{314}+E^{235},&
E^{6}_{2} ~ &=& ~ E^{4}_{~2}+E^{5}_{~1},&
E^{6}_{3} ~ &=& ~ -(E^{*}_{256}+E^{*}_{164}),\n
E^{7}_{1} ~ &=& ~ i(E^{125}-E^{316}),&
E^{7}_{2} ~ &=& ~ i(E^{5}_{~3}-E^{6}_{~2}),&
E^{7}_{3} ~ &=& ~ -i(E^{*}_{364}-E^{*}_{245}),\n
E^{8}_{1} ~ &=& ~ i(E^{236}-E^{124}),&
E^{8}_{2} ~ &=& ~ i(E^{6}_{~1}-E^{4}_{~3}),&
E^{8}_{3} ~ &=& ~ -i(E^{*}_{145}-E^{*}_{356}),\n
E^{9}_{1} ~ &=& ~ i(E^{314}-E^{235}),&
E^{9}_{2} ~ &=& ~ i(E^{4}_{~2}-E^{5}_{~1}),&
E^{9}_{3} ~ &=& ~ -i(E^{*}_{256}-E^{*}_{164}).
\end{array}
\n
\label{Egenerators}
 \end{eqnarray}

\end{enumerate}

One can verify that the 78 generators (\ref{Ehatgenerators}),
(\ref{Tgenerators}),(\ref{Estargenerators}) and (\ref{Egenerators}) 
form a closed algebra, which is a real form of $E_6$ by construction. 
By examining the Killing form, it can be easily seen that 
the $SL(3,\CC)$ generators (\ref{Tgenerators}) 
as well as the $E_I^{*i}$ 
and $E_i^I$ generators (\ref{Estargenerators}) (\ref{Egenerators}) 
consist of the same number of positive and negative generators.
This leaves the $SL(3,\RR)$ (\ref{Ehatgenerators}), which is $A_{2(+2)}$. 
Thus we see that the whole algebra is $E_{6(+2)}$.

One can also show that it satisfies the commutation relations of the form 
(\ref{commutationrelations}), 
whose actual values of the structure constants are 
 \begin{eqnarray}
C^{123} ~ &=& ~ +\sqrt{2},\nonumber\\
C^{144} ~ = ~ C^{177} ~=~ C^{255} ~=~ C^{288} ~=~ C^{366} ~=~ C^{399} ~&=&~ -\sqrt{2},\nonumber\\
C^{456}  ~&=&~ +1,\nonumber\\
C^{489} ~=~ C^{579} ~ = ~ C^{678} ~ &=& ~ -1,
 \end{eqnarray}
and
 \begin{eqnarray}
D^{1~3}_{~5}~=~D^{1~2}_{~6}~=~D^{2~6}_{~4}~=~
D^{2~4}_{~6}~=~D^{3~8}_{~4}~=~D^{3~7}_{~5}~&=&~\nonumber\\
D^{4~8}_{~2}~=~D^{4~6}_{~3}~=~D^{5~7}_{~1}~=~
D^{5~3}_{~3}~=~D^{6~4}_{~1}~=~D^{6~2}_{~2}~&=&~+\sqrt{2},\nonumber\\
D^{4~4}_{~5}~=~D^{4~7}_{~6}~=~D^{5~2}_{~4}~=~
D^{5~8}_{~6}~=~D^{6~3}_{~4}~&=&~\nonumber\\
D^{6~6}_{~5}~=~D^{6~1}_{~6}~=~D^{6~5}_{~6}~=~
D^{9~1}_{~9}~=~D^{9~5}_{~9}~&=&~+1,\nonumber\\
D^{4~1}_{~4}~=~D^{5~5}_{~5}~=~D^{7~1}_{~7}~=~
D^{7~4}_{~8}~=~D^{7~7}_{~9}~&=&~\nonumber\\
D^{8~2}_{~7}~=~D^{8~5}_{~8}~=~D^{8~8}_{~9}~=~
D^{9~3}_{~7}~=~D^{9~6}_{~8}~&=&~-1,\nonumber\\
D^{1~1}_{~1}~=~D^{2~5}_{~2}~&=&~+2,\nonumber\\
D^{3~1}_{~3}~=~D^{3~5}_{~3}~&=&~-2,\nonumber\\
\nonumber\\
D^{1~10}_{~9}~=~D^{2~14}_{~7}~=~D^{3~15}_{~8}~=~
D^{7~14}_{~3}~=~D^{8~15}_{~1}~=~D^{9~10}_{~2}~&=&~+\sqrt{2},\nonumber\\
D^{1~11}_{~8}~=~D^{2~12}_{~9}~=~D^{3~16}_{~7}~=~
D^{7~16}_{~2}~=~D^{8~11}_{~3}~=~D^{9~12}_{~1}~&=&~-\sqrt{2},\nonumber\\
D^{4~12}_{~8}~=~D^{5~13}_{~8}~=~D^{5~16}_{~9}~=~
D^{6~11}_{~7}~=~D^{6~13}_{~9}~&=&~\nonumber\\
D^{7~9}_{~4}~=~D^{7~12}_{~5}~=~D^{8~16}_{~6}~=~
D^{9~11}_{~4}~=~D^{9~9}_{~6}~&=&~+1,\nonumber\\
D^{4~9}_{~7}~=~D^{4~15}_{~9}~=~D^{5~10}_{~7}~=~
D^{6~14}_{~8}~=~D^{6~9}_{~9}~&=&~\nonumber\\
D^{7~15}_{~6}~=~D^{8~10}_{~4}~=~D^{8~13}_{~5}~=~
D^{9~14}_{~5}~=~D^{9~13}_{~6}~&=&~-1,\nonumber\\
D^{5~9}_{~8}~=~D^{7~13}_{~4}~&=&~+2,\nonumber\\
D^{4~13}_{~7}~=~D^{8~9}_{~5}~&=&~-2,
 \end{eqnarray}
otherwise 0. 
Thus we have confirmed that $E_{6(+2)}=\mbox{qConf}({\rm J}_3^{\CCsub})$
and $F_{4(+4)}=\mbox{qConf}({\rm J}_3^{\RRsub})$ share the common 
 algebraic structure in terms of the $SL(3,\RR)\times
\mbox{Str}_0({\rm J}_3^{\AAsub})$ decomposition, whose coset 
necessarily leads, as we proved in section 2, to the general form 
of the three-dimensional sigma model of a dimensionally reduced 
magical supergravity. This means that
 the second, ${\rm J}_3^{\CCsub}$ 
magical supergravity also 
allows the algebraic interpretation of the structure constants and 
scalar field functions as was done in the first, ${\rm J}_3^{\RRsub}$  
magical supergravity.  We naturally expect that the remaining 
magical supereravities also possess such a structure. We hope to report 
on this problem elsewhere.

\section*{Acknowledgments} 
We would like to thank Naoto Kan, 
Hideo Kodama and Masato Nozawa for discussions. %
The work of S.~M. is supported by 
Grant-in-Aid
for Scientific Research  
(C) \#16K05337 
and 
(A) \#26247042
from
the Ministry of Education, Culture, Sports, Science
and Technology of Japan.

\end{document}